# Intelligent EC Rearview Mirror: Enhancing Driver Safety with Dynamic Glare Mitigation via Cloud Edge Collaboration

Junyi Yang[1], Zefei Xu[2], Huayi Lai[3], Hongjian Chen[2], Sifan Kong[1], Yutong Wu[1], Huan Yang[1*]

*Abstract*— Sudden glare from trailing vehicles significantly increases driving safety risks. Existing anti-glare technologies such as electronic, manually-adjusted, and electrochromic rearview mirrors, are expensive and lack effective adaptability in different lighting conditions. To address these issues, our research introduces an intelligent rearview mirror system utilizing novel all-liquid electrochromic technology. This system integrates IoT with ensemble and federated learning within a cloud edge collaboration framework, dynamically controlling voltage to effectively eliminate glare and maintain clear visibility. Utilizing an ensemble learning model, it automatically adjusts mirror transmittance based on light intensity, achieving a low RMSE of 0.109 on the test set. Furthermore, the system leverages federated learning for distributed data training across devices, which enhances privacy and updates the cloud model continuously. Distinct from conventional methods, our experiment utilizes the Schmidt-Clausen and Bindels de Boer 9-point scale with TOPSIS for comprehensive evaluation of rearview mirror glare. Designed to be convenient and cost-effective, this system demonstrates how IoT and AI can significantly enhance rearview mirror anti-glare performance.

## I. INTRODUCTION

Road traffic safety, a pressing global issue, hinges on the driver's rapid and precise interpretation of visual information for immediate decision-making [1]. Notably, the images in rearview mirrors serve as vital visual cues for drivers. However, abrupt changes in nighttime lighting and intense light refraction can cause glare that severely hampers visual clarity and decision-making in the rearview mirror, presenting significant safety risks [2]. Mitigating this glare is crucial for improving traffic safety and intelligence. Despite advancements in anti-glare technologies, their high costs and operational challenges hinder widespread adoption. For instance, electronic rearview mirrors are costly, and traditional anti-glare mirrors require manual adjustments [3]. Electrochromic (EC) technology, which intelligently adjusts transmittance, has proven effective in smart window. However, current EC mirrors still apply the same voltage when detecting different intensities of glare, limiting their effectiveness. Although smart control technology is well-established in smart windows, its application in enhancing EC rearview mirrors remains underdeveloped [4].

This study introduces an intelligent rearview anti-glare mirror utilizing all-liquid EC materials alongside ensemble and federated learning within a cloud edge collaboration framework. The system dynamically adjusts the output voltage to change the transmittance of the rearview mirrors and optimize the anti-glare function, thus enhancing the driving experience and safety. Furthermore, this advancement addresses existing technological shortcomings, particularly in adaptive response, underscoring the potential of integrating Internet of Things (IoT) and artificial intelligence (AI) into rearview mirror system. The system integrates three main modules: Rearview Mirrors Monitoring & Control Based on IoT & Edge Computing (IoT Frontend) for mirror adjustments, Ensemble-Learning-Based Voltage Regulation Level Predictor (EL-Based Predictor) utilizing XGBoost and MLP in a federated learning model, and Human-Machine Interface Facilitating Multi-Device Control (HMI), accessible via a WeChat mini-program. The key contributions of this study are as follows:

1) To the best of our knowledge, this is the first development of an intelligent rearview mirror system integrating EC technology, AI algorithms, and IoT. The system incorporates a novel all-liquid EC device, uniquely developed by our team [7].

2) Traditional glare measurement methods require specialized equipment and experiments, yet studies on rearview mirror glare are limited. Our research uses the Schmidt-Clausen and Bindels de Boer 9-point scale with TOPSIS to quantify subjective glare assessments, enhancing evaluation accuracy and validity.

3) Our system employs an ensemble learning model, combining XGBoost and MLP algorithms to construct the base model. It also incorporates federated learning within a cloud edge collaboration framework to incrementally update this base model in the cloud. These updates are based on manual adjustments made by users, allowing for continuous, incremental improvement of model performance.

The remainder of this article is organized as follows. Section II describes the related work in detail, Section III describes the design architecture of the system, Section IV analyses the performance of the system, and Section V concludes with future directions.

## II. RELATIVE WORK

### A. Application of Electrochromic (EC) Devices

Smart windows, adjusting indoor temperatures by altering light transmission properties, have become a significant area of research that is expected to help reduce cooling energy demands. EC devices are particularly notable in smart window technologies [5]: These devices change color under different electric currents and are categorized into organic and inorganic materials, and into solid, liquid, and gel states [6]. Known for high light transmission modulation, rapid cycling, and efficient coloration, EC devices have advanced significantly. Notably, Kong et al., developed an all-liquid EC device recently, using a blend of ammonium paratungstate and iron chloride [7]. This device, simple and cost-efficient, with superior performance, serves as an ideal prototype for our investigations.

*Corresponding author: Huan Yang (20212034008@m.scnu.edu.cn)

[1]J. Yang, S. Kong, Y. Wu and H. Yang are with the School of Software, South China Normal University, Foshan 528225, China.

[2]Z. Xu and H. Chen are with the School of Electronics and Information Engineering, South China Normal University, Foshan 528225, China.

[3]H. Lai is with the Aberdeen Institute of Data Science and Artificial Intelligence, South China Normal University, Foshan 528225, China.

*B. The IoT, Machine Learning, and EC Devices*

The rapid advancements in machine learning and IoT technologies have started to impact the field of EC devices, enhancing the development of new EC technologies. For example, Kong et al. utilized MLP to assist in developing EC devices [8], while Brandon et al. explored the correlation between the EC properties of $WO^3$ films and sputtering parameters using machine learning [9]. IoT technology has also been employed to analyze indoor environmental data for smart windows, as demonstrated by Khatibi et al. and Azhar et al., who designed and presented IoT-based systems for smart window adjustments [10-11]. However, despite these advancements, challenges remain regarding the design of systems based on EC devices that exhibit adequate ambient intelligence and minimize user intervention.

### III. ARCHITECTURE

The architectural framework of the rearview mirror system employed in this study comprises three main modules: A) IoT Frontend, B) EL-Based Predictor, and C) HMI. Each module is dedicated to specific functions within the system. The IoT Frontend integrates an array of sensors and EC devices to accurately detect light intensity and implement precise adjustments to the EC effect, leveraging voltage predictions provided by the EL-Based Predictor. The EL-Based Predictor accurately predicts voltage levels, effectively eliminating glare, and maintaining clear visibility. Additionally, the HMI module enables real-time control over the rearview mirrors, accommodating both automated and manual adjustments. Furthermore, certain modules undergo further subdivision into sub-modules, as elaborated subsequently.

*A. Rearview Mirrors Monitoring & Control Based on IoT & Edge computing (IoT Frontend)*

Within the system, the IoT Frontend acts as the foundational module, responsible for receiving external light intensity data and controlling the hardware components via edge computing. The Raspberry Pi 4B+, chosen for its robust processing capabilities, serves as the microcontroller unit at the core of this module. It features a 1.5GHz ARM Cortex-A72 processor with options for 1GB, 2GB, or 4GB of LPDDR4-3200 SDRAM and supports Gigabit Ethernet, dual-band Wi-Fi, and Bluetooth 5.0 connectivity. For hardware connectivity, the Raspberry Pi's 40-pin connector interlinks various sensor and control chips. The circuit diagram of the IoT Frontend is illustrated in Fig. 2. IoT Frontend is further divided into two sub-modules: Sensor Data Acquisition and Linear Regulated Voltage Control.

1) Sensor Data Acquisition module: This sub-module utilizes two photoresistors to convert light intensity into analog voltage signals. These are then converted into digital signals by ADC chip, facilitating precise and timely data processing. These values are then transmitted to the Raspberry Pi via the SPI protocol for further processing.

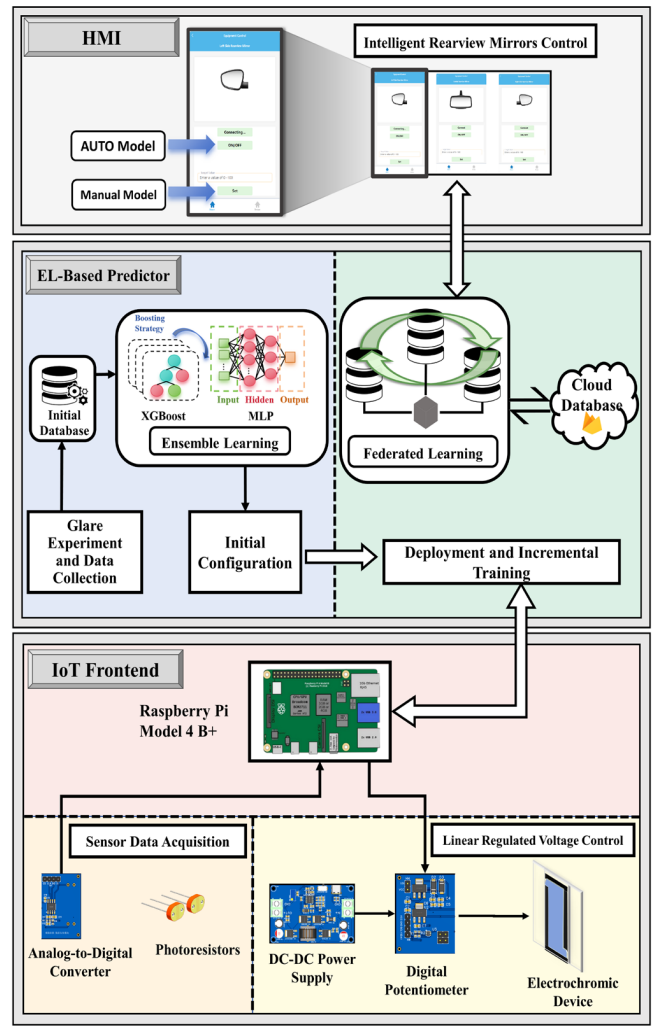

Figure 1. The architecture of the rearview mirror system in this study

a) The GL5506 photoresistors are used to distinguish between incident and ambient light intensities. Employing two units, one sensor captures incident light, while the other captures background illumination. These variations in light intensity are converted into analog voltage signals, enabling precise differentiation of lighting conditions by the circuit.

b) The TLC0832 ADC chip, operating at a clock frequency of up to 250kHz, converts analog voltages into digital signals. It can simultaneously process voltages from both photoresistors with a precision up to 0.01V, thus exceeding the system's accuracy requirements.

2) Linear Regulated Voltage Control: In this design, the Raspberry Pi utilizes the Inter-Integrated Circuit (IIC) protocol to send precise control signals to a digital potentiometer, forming an integral part of the linear voltage regulator circuit. This precise regulation is essential for controlling the transmittance of EC devices, ensuring change in response to changes in the vehicle's ambient lighting.

a) The MCP4017 digital potentiometer, with its 128 taps, facilitates voltage adjustment across 128 distinct levels and enables precise adjustment of output voltages by receiving control signals from the Raspberry Pi for regulation.

b) Experimental findings indicated that the 5V GPIO pins on the Raspberry Pi were inadequate for the power demands of the EC device. To address this, we implemented a dedicated 5V power module paired with an AMS1117 voltage regulator. This setup provided a stable voltage output ranging from 1.49V to 3.79V, covering the required operational voltage range for EC device.

c) All-liquid EC device was incorporated by the system, featuring EC technology, electrolytes, and an ion storage layer. A transparent conductive layer serves as both the shell and substrate, enabling dynamic color changes via redox reactions. The device employs ammonium metatungstate (0.1 g/mol) and ferrous chloride (0.15 g/mol) as EC materials, blended with heavy water and deionized water in a 4:1 ratio. Notably, this EC device exhibits a high reaction rate, extensive light transmittance range, robust stability, and straightforward preparation.

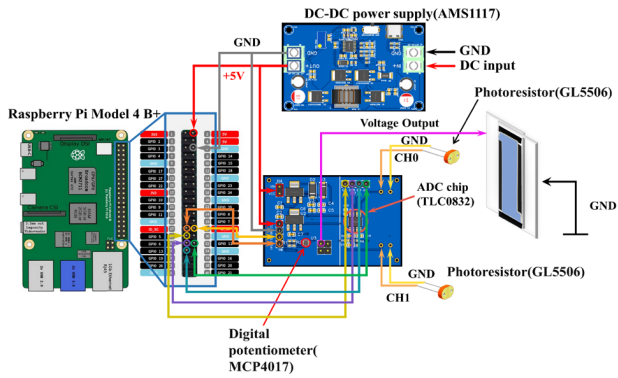

Figure 2. The circuit diagram of the IoT Frontend

### B. Ensemble-Learning-Based Voltage Regulation Level Predictor (EL-Based Predictor)

The EL-Based Predictor dynamically predict the applied voltage of EC devices using Machine Learning (ML) algorithms to mitigate glare from trailing vehicles' headlights. This module is structured into three integral sub-modules: firstly, the glare experiment and data collection module, designed for glare assessment and data acquisition; secondly, the initial configuration module, which utilizes an ensemble learning framework to establish the model's initial settings; and thirdly, the deployment and incremental training module, aimed at the model's continual enhancement and deployment through federated learning between the cloud and edge devices. This organized approach ensures each module contributes effectively to the overarching functionality of the EL-Based Predictor.

#### 1) Glare Experiment and Data Collection Module

Many existing glare measurement methods need specialized equipment or precise experiments, and none specifically evaluate rearview mirror glare. Therefore, this study utilizes the Schmidt-Clausen and Bindels de Boer 9-point glare rating scale, which is specifically designed to precisely assess discomfort glare from vehicle headlights. These evaluations directly inform the optimization of the voltage adjustment strategy for EC rearview mirrors. If a participant's glare rating exceeds the 'Acceptable' (rating 7) threshold, researchers gradually increase the voltage until the glare is reduced to 'Acceptable' levels. This method is designed to mitigate the negative impact of glare on drivers, while simultaneously maintaining the clarity of the rearview mirror. To ensure sample diversity and representativeness, ten drivers with varied genders and vision conditions were recruited via snowball and quota sampling methods. Acknowledging that incident light intensity and contrast significantly influence glare [3], our experiment simulates various lighting conditions across day and night to determine optimal rearview mirror visibility without glare interference. The procedure of the experiment is illustrated in Fig. 3.

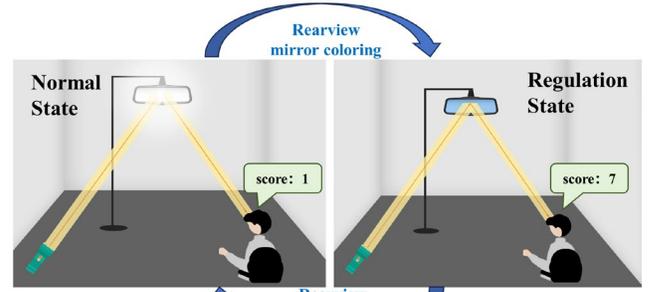

Figure 3. Glare level experiment scenario

To systematically translate subjective glare discomfort ratings into quantifiable metrics, the Technique for Order Preference by Similarity to Ideal Solution (TOPSIS) was employed. The methodology commenced with the computation of the contrast ratio, achieved by subtracting front sensor readings from those at the rear. This calculated contrast, in conjunction with incident light intensity data, was subsequently analyzed through TOPSIS to derive objective scores. Further generalization and grading procedures were conducted. The outcomes of this quantitative analysis are meticulously documented in Table I, aligning the Schmidt-Clausen and Bindels scales with the corresponding TOPSIS scores. This categorization spans from 'Unbearable' to 'Noticeable,' facilitating the evaluation of glare mitigation strategies and enabling precise calibration of electrochromic mirror settings. These objective scores quantify the prevailing levels of rearview mirror glare, thereby offering potential enhancements to driving conditions under diverse lighting scenarios.

TABLE I. THE SCHMIDT-CLAUSEN AND BINDELS SCALE AND TOPSIS SCORE

| Assessment | Glare rating W | TOPSIS score |
|---|---|---|
| Unbearable | 1 | [1,0.7671) |
|  | 2 | [0.7671,0.6576) |
| Disturbing | 3 | [0.6576,0.5207) |
|  | 4 | [0.5207,0.3015) |
| Just admissible | 5 | [0.3015,0.2192) |
|  | 6 | [0.2192,0.1644) |
| Acceptable | 7 | [0.1644,0.0548) |
|  | 8 | [0.0548,0) |
| Noticeable | 9 | 0 |

#### 2) Initial Configuration Module

This module is based on experimental data and employs an ensemble learning framework using XGBoost and MLP as base learners in a stacking generative model, with Ridge Regression acting as the meta-model to improve prediction accuracy. It consists of four main components: Stacked Generation, XGBoost, MLP, and Hyperparameter Tuning.

Subsequently, the trained model is transferred to the cloud, where it is further refined and train, enhancing overall model deployment and scalability.

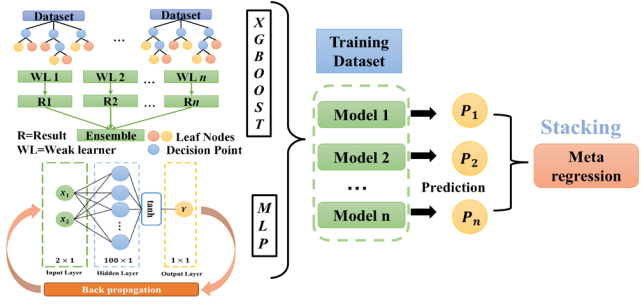

Figure 4. The Stacking generation process

*a) Stacking Generation*

Stacking generation enhances prediction accuracy by combining multiple regression models through a meta-regressor, effectively leveraging the strengths of diverse algorithms. In this method, each base algorithm is trained on the full dataset, and their outputs serve as meta-features, which the meta-regressor uses to refine final predictions. Specifically, our system employs Ridge Regression as the meta-regressor, with XGBoost and MLP serving as the base models. The stacking generation process is shown in Fig. 4.

*b) XGBoost*

XGBoost leverages decision trees to create numerous weak learners, enhancing them into a robust ensemble through gradient boosting. Its regularization prevents overfitting and ensures adaptability across various intensities of glare. See the objective function of XGBoost below.

$$X_{Obj} = \sum_{i=1}^{n} l(y, \hat{y}) + \sum_{k=1}^{k} \Omega(f) \quad (1)$$
$$\Omega(f) = \gamma T + \lambda \frac{1}{2} \sum_{j=1}^{T} w_j^2$$

The formula above outlines the XGBoost algorithm's objective function, $X_{Obj}$ combining the loss function $l(y, \hat{y})$ —the difference between actual and predicted regulation voltages—and the regularization term $\Omega(f)$ that penalizes tree complexity. $T$ counts leaf nodes, while $\gamma$ and $\lambda$ act as penalty factors for complexity and leaf weights.

*c) MLP*

The MLP utilized in our study comprises four layers: an input layer, two hidden layers, and an output layer. Leveraging forward computation and backpropagation, the MLP dynamically adapts to changing glare intensity, as expressed mathematically:

$$z^i = R(W^i z^{i-1} + b^i) \quad (2)$$

In this formula, $z^i$ is the output of the $i$ layer; $W^i$ and $b^i$ are the weights and biases between the $i$ layer and the $i-1$ layer respectively; $R$ is the activation function.

*d) Hyperparameter Tuning*

Hyperparameter optimization enhances machine learning model performance. Grid Search, implemented through Scikit-learn, identifies optimal settings for MLP and XGBoost algorithms. Stacking optimized models further improves regression outcomes, with key parameters detailed in Table II. All other settings are kept at default values.

TABLE II. ESSENTIAL OPTIMAL PARAMETERS

| Algorithm | Parameter | Value |
| --- | --- | --- |
| Stacking | final_estimator | Ridge() |
| XGBoost | learning_rate | 0.1 |
| | n_estimators | 50 |
| MLP | activation | tanh |
| | alpha | 0.001 |
| | hidden_layer_sizes | (100,) |

*3) Deployment and Incremental Training Module*

Engineered for incremental model training, this module employs federated learning to effectively train models across distributed devices. Data is processed locally on each device, and only model parameter updates are shared, which enhances privacy by minimizing data exposure. Each rearview mirror device functions as an independent node, responsible for data collection, processing, and local training using ensemble learning model.

After training, these devices upload their model parameters to a Firebase cloud database for aggregation. The aggregation at the cloud server incorporates both the frequency of manual mode usage and the recency of data from each device to formulate the global model parameters using the formula:

$$\Theta = \frac{\sum_{i=1}^{N} \lambda^{t_i} u_i \theta_i}{\sum_{i=1}^{N} \lambda^{t_i} u_i} \quad (3)$$

In this formula, $\theta_i$ represents model parameters from node $i$, $u_i$ is the manual mode usage count by node $i$, $\lambda$ acts as a decay factor, and $t_i$ is the time since the last update, prioritizing recent data. $\Theta$ denotes the aggregated global model parameters. To improve model accuracy, an optimization step corrects for prediction errors:

$$\Theta_{optimized} = \Theta + \alpha \left( \frac{\sum_{i=1}^{N} \overline{e}_i \theta_i}{\sum_{i=1}^{N} \overline{e}_i} \right) \quad (4)$$

This step uses $\Theta$, the aggregated global model parameters, and adjusts them based on aggregated error metrics $\overline{e}_i$ computed locally at each node. Here, $\alpha$ is a tuning parameter that determines the extent of correction applied, enabling the model to adapt based on summarized error data rather than detailed individual errors. $\Theta_{optimized}$

reflects the refined global model parameters. A schematic diagram of the architecture can be seen in Fig. 5.

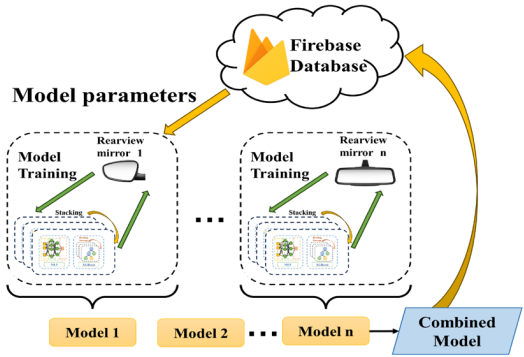

Figure 5. The structure of Federated learning model

### C. Human-Machine Interface Facilitating Multi-Device Control (HMI)

The HMI improves IoT device management and control with a visual platform, using WebSocket Secure (WSS) for encrypted communication and Message Queuing Telemetry Transport (MQTT) for lightweight messaging, ensuring secure, stable communication. It features real-time command reception and better device monitoring, consisting of two sub-modules: Intelligent Rearview Mirrors Management and Control. The GUI for HMI is shown in Fig. 6.

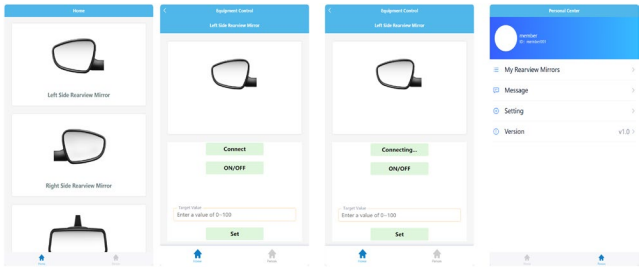

Figure 6. The GUI for HMI

#### 1) Intelligent Rearview Mirrors Management Module

This module manages multiple rearview mirror devices simultaneously through a WeChat mini-program, utilizing the WSS protocol to ensure secure, real-time updates of device statuses.

#### 2) Intelligent Rearview Mirrors Control Module

This module provides both manual and automatic voltage control options, enabling users to adjust mirror voltage either through a user-friendly interface or automatically, based on sensor data, to optimize visual conditions.

## IV. PERFORMANCE ANALYSIS

### A. Prototype Implementation

The system uses two photoresistors to collect light intensity data, converting these readings into digital signals with the TLC0832 ADC chip. These signals are then transmitted to the Raspberry Pi for processing via the SPI protocol, ensuring efficient data handling. Additionally, the Raspberry Pi manages the voltage regulator circuit using the IIC protocol, enabling precise voltage control of the EC device and effective glare suppression. The hardware connection schematic of the system is shown in Fig. 7.

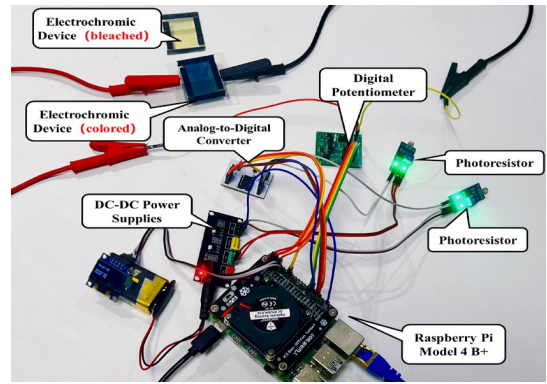

Figure 7. Physical drawing of the EC Rearview mirror in this study

### B. Performance for EC Rearview Mirror

The EC mirrors developed in this study have a large modulation amplitude and fast response time to optimize visibility under different glare intensities. Key performance indicators are presented in Table III. For a detailed performance, refer to the study by co- author Kong [8].

TABLE III. PERFORMANCE OF EC REARVIEW MIRROR

| Performance Indicator | Value |
|---|---|
| Switching voltage | $\leq 5\ V$ |
| Switching time | $\leq 10\ s$ |
| Stable current | $\leq 100\ mA$ |
| Rated voltage | $\leq 2\ V$ |
| Lifetime | 25000-50000cycles <br> 5-10year |
| Transmittance | 70-80%(bleached) <br> 6-10%(colored) |

### C. Performance Analysis of EL-Based Predictor

Initially, in the Initial Configuration module, this study conducted a comparative analysis of various machine learning algorithms implemented on the same dataset. The results of this study demonstrate that the stacking ensemble learning algorithm outperforms other algorithms, achieving superior accuracy and fit as indicated by its higher coefficient of determination ($R^2$) and lower Root Mean Square Error (RMSE). Specifically, in the test set, the Stacking algorithm achieved an $R^2$ of 0.9487 and an RMSE of 0.109. Fig. 8 illustrates the comparison of $R^2$ and RMSE across six machine learning algorithms.

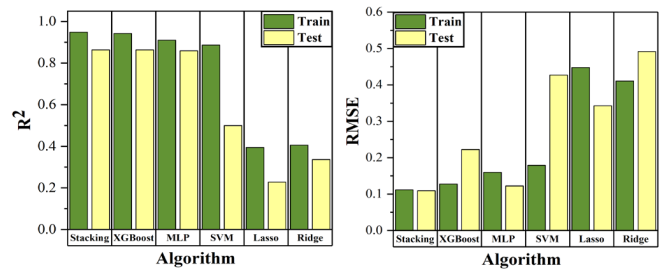

Figure 8. Comparison of $R^2$ and RMSE of 6 machine learning algorithms

This study examines federated learning's impact on model performance through simulations across five nodes over ten rounds. The findings, illustrated in Fig. 9, reveal consistent improvements in $R^2$ values for both training and test datasets, highlighting federated learning's potential to enhance algorithm training efficiency.

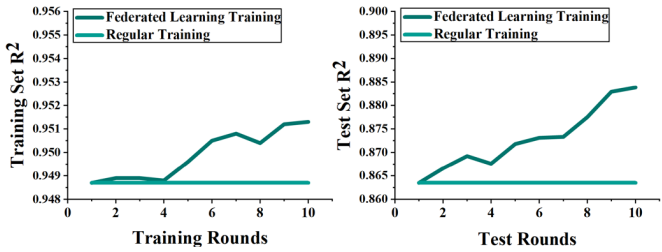

Figure 9. Algorithm Performance: before and after federated learning

To analyze the scalability and responsiveness of federated learning systems, the study methodically increased the number of nodes and conducted ten rounds of federated learning at each increment. Subsequently, it calculated the average CPU usage and aggregation time to quantitatively evaluate the impact of nodes proliferation on system performance. Detailed variations in CPU usage and aggregation time are systematically documented in Fig. 10.

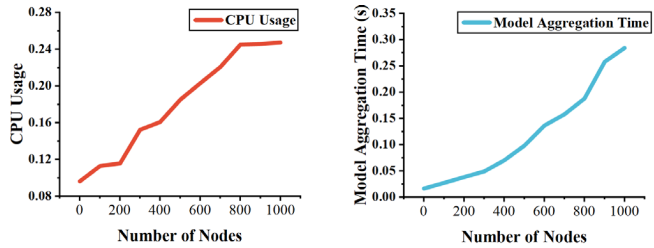

Figure 10. CPU usage and aggregation time trends by the number of nodes

### D. Evaluating the Glare Reduction Effectiveness of the Developed EC Mirrors

In order to evaluate the effectiveness of EC mirrors in reducing glare in this paper, we conducted an experiment with 10 participants in six glare conditions based on the data presented in Table I. The mirrors automatically adjusted the voltage according to changing light conditions via the EL-Based Predictor. Participants experienced each scenario and recorded the light intensity and contrast before and after activating the mirrors. The results were converted to TOPSIS scores and averaged as shown in Fig.11, demonstrating a reduction in glare discomfort and confirming the mirrors' ability to improve driving safety.

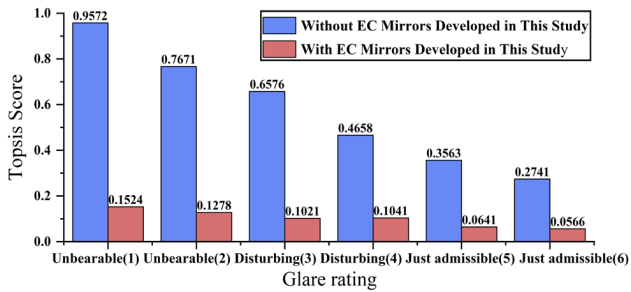

Figure 11. Analysis of TOPSIS scores for glare reduction with EC mirrors

## V. CONCLUSION

In summary, this study introduces an intelligent rearview anti-glare mirror system that combines all-liquid EC materials with edge computing, IoT, ensemble learning, and federated learning. Such integration significantly enhances night-time driving safety by dynamically adjusting mirror transmittance to effectively minimize glare, thereby preserving image clarity. Concurrently, the performance of the developed ensemble algorithm was benchmarked against existing ML algorithms, enhanced through federated learning, and assessed for scalability and responsiveness. The effectiveness of the EC mirrors developed in this study for glare reduction was evaluated using the TOPSIS.

The system base model effectively employs the ensemble learning approach, but ignores the long-term performance degradation of EC devices. Consequently, future research will integrate time-series models to predict and enhance the performance and regulation accuracy of EC devices. Additionally, the dataset will expand to include varied glare measurements. Scalability and responsiveness analyses of federated learning indicate that increasing nodes numbers strain system capacity, necessitating dynamic adaptations for peak performance. Furthermore, the smart control algorithms developed here may enhance other anti-glare mirrors and, integrated into telematics networks, could enrich the driving experience with more responsive vehicle interactions.